\documentclass[preprint,12pt]{elsarticle}

\usepackage{amssymb}
\usepackage{amsmath}
\usepackage{mathtools}

\journal{Nuclear Physics B}

\begin{document}

\begin{frontmatter}



\title{Uncertainty Quantification in Computational Fluid Dynamics: Physics and Machine Learning Based Approaches}


\author[inst1]{Minghan Chu}

\affiliation[inst1]{organization={Queen's University},
            city={Kingston},
            state={Ontario},
            country={Canada}}

\begin{abstract}
Turbulent flows are important in many problems for engineering design and scientific analysis. At present and for the foreseeable future, computational studies of turbulent flows rely on turbulence models, both in Reynolds Averaged Navier Stokes (RANS) based modeling and Sub-Grid Scale (SGS) models in Large Eddy Simulations (LES). Turbulence model based simulations suffer from many different sources of prediction uncertainties. For example simplifications and approximations used to make these turbulence models computationally tractable and inexpensive lead to predictive uncertainty. In safety critical applications of engineering design we need reliable estimates of these predictive uncertainties. This article focuses on Uncertainty Quantification (UQ) for Computational Fluid Dynamics (CFD) simulations. We review recent advances in estimating different components of uncertainty, including aleatoric, numerical and epistemic. We elaborate upon the use of Machine Learning (ML) algorithms for estimating these uncertainties. Most critically, we elaborate upon seminal limitations in these approaches. These range from realizability constraints on the Eigenspace Perturbation Method (EPM) to the requirement for Monte Carlo (MC) approaches for mixed uncertainties. Based on this analysis we highlight central questions that need to be addressed and advocate focused steps to redress these limitations. 
\end{abstract}




\end{frontmatter}


\section{Introduction }
\label{sec:sample1}

Turbulent flow has been extensively studied using computational fluid dynamics (CFD) simulations since turbulent flow regime is so frequently encountered in both academic and engineering applications. The high-fidelity simulation of the Direct Numerical Simulation (DNS) requires a sufficiently fine mesh to resolve the smallest Kolmogorov length scale of turbulent motion, which requires tremendous amount of computational resources, and hence usually prohibited in engineering applications. At the current state of computational power we can only execute DNS simulations for small sections of an aircraft wing at high Reynolds numbers. Even though large-eddy simulation (LES) has reduced the computational overheads by only resolving the large-scale eddies with the small-scale eddies being modelled. However, for extremely high Reynolds number and complex geometry flows LES is still restricted to academic sttudies. As the result, Reynolds-averaged Navier-Stokes (RANS) based turbulence models have significantly reduced the computational overheads by modeling all scales of turbulent motion. Unlike DNS and LES which are dedicated to representing the true physics of turbulent flows, RANS approach uses simplifying modeling assumptions to describe both lower order and higher order quantities. This makes RANS still remain the most widely used CFD method in engineering applications; however, simplifying assumptions also introduce sources uncertainties during simulation. 


Sources of uncertainty can be in general classified as aleatory and epistemic \cite{smith2013uncertainty}. An appropriate description of uncertainty, e.g., probabilistic distribution, is needed for uncertainty quantification, which essentially aims to estimate the effect of propagated uncertainties on the output predictions and then reduce uncertainties. Aleatory uncertainties are typically introduced in the inputs (initial and boundary conditions), and the natural variability (imprecision) of a system \cite{duraisamy2019turbulence}. Sources of aleatory uncertainties include errors in measurement for geometry and material properties, differences in the initial or boundary conditions between simulations and reality, etc. Therefore aleatory uncertainties are irreducible/unbiased, inherently stochastic (typically described as probability), and can result in a proliferation of complexity in any simulation of a real-world system. On the other hand, aleatory uncertainties can be reduced by reconstructing the stochastic terms to achieve more accurate initial and boundary conditions (priors). The variance inherent in aleatory uncertainties can be propagated throughout the simulation. Numerous UQ strategies have been developed to quantify and reduce aleatory uncertainties. The statistical inference strategies: statistical (inference) strategies like Stochastic Collocation \cite{mathelin2003stochastic} and Polynomial Chaos \cite{najm2009uncertainty} approaches inspired by Bayes theorem assign the posterior probability distributions to model parameters through a calibration process. This includes representing model coefficients as random variables \cite{loeven2008airfoil,ahlfeld2017single}, representing flow domain as a stochastic field \cite{dow2015implications,doostan2016bi}, and representing the initial and boundary conditions as stochastic \cite{pecnik2011assessment}, etc.

On the other hand, epistemic uncertainties are due to model inadequacy, i.e., inherent inability to represent the physics of turbulence in models. This type of uncertainty is reducible/biased and will still remain even if aleatory uncertainties are diminished. Epistemic uncertainty includes uncertainty associated with the coefficients of a turbulence model and the structural uncertainty associated with the limitations of the model expression. It should be noted that both aleatory and epistemic uncertainties can be represented by probability distribution in the Bayesian framework, as long as sufficient prior information is available to construct one \cite{najm2009uncertainty}. According to Duraisamy \textit{et al.} \cite{duraisamy2017status} structural uncertainty can be a dominant source of uncertainty in everyday engineering modeling of turbulent flows. Epistemic uncertainties can be mitigated by employing high-fidelity models, which, again, are usually accompanied by unaffordable computational expenses. Therefore, it is worth using a RANS-based eddy-viscosity model to retain the lower computational expense while taking advantage of quantifying the structural uncertainties. Currently the Eigenspace Perturbation Method (EPM) \cite{emory2013modeling,iaccarino2017eigenspace} is the only physics-based approach that is available to estimate the uncertainty due to turbulence models. In the EPM we perturb the spectral decomposition components of the predicted Reynolds stress tensor to estimate the sensitivity of the predictions due to the structural uncertainties. Since no high demand on an extremely fine mesh and sufficient a \textit{priori} data are required, the EPM can estimate the propagated effect of uncertainties on predictions at a very low cost and is applicable for general turbulent flows. This method has been successfully applied across different fields of engineering including the design of urban canopies\cite{gorle2019epistemic}, aerospace design and analysis of  \cite{mishra2019uncertainty, mishra2017rans, mishra2019estimating, mishra2017uncertainty}, application to design under uncertainty (DUU) \cite{demir2023robust, cook2019optimization, mishra2020design, righi2023uncertainties, alonso2017scalable}, virtual certification of aircraft designs \cite{mukhopadhaya2020multi, nigam2021toolset, mishra2019linear}, the design of wind farms, etc. 


Machine Learning (ML) based models are being increasingly applied to fluid dynamics and turbulence applications\cite{duraisamy2019turbulence, ihme2022combustion, brunton2020machine, chung2021data}. Numerous investigators have used data to develop functions that can predict the discrepancy in turbulence model predictions \cite{xiao2016quantifying,wu2018physics,heyse2021estimating,heyse2021data,zeng2022adaptive}. The focus of these studies has been turbulence model form or epistemic uncertainty estimation. But many of these ML models have a tendency to overfit. This limits the generalizability of the ML models and the trained models end up being accurate only for flows that they are trained on. The complex nature of the machine learning models also requires large amounts of relevant data for training them. In engineering design this is not always possible especially when new designs are being considered. The complex machine learning models are also prone to being black box models where their inner working is not well understood. This limits our ability for Verification and Validation of such models leading to issues of trust and with adoption of these ML models in engineering applications \cite{chung2022interpretable}.  While data is specific to the training cases it was collected from, physics principles are universal and apply to all flows. So incorporating physics knowledge in these ML models for uncertainty quantification is an essential requirement.

In addition to aleatory and epistemic uncertainties, numerical or algorithmic uncertainty arises due to discretization or schemes to solve PDEs. The common method to quantify the discretization uncertainty is called Grid Convergence Index (GCI) of Roache \cite{roache1993method,roache1997quantification,roache1998verification} and Richardson Extrapolation (or $h^2$ extrapolation) \cite{richardson1927viii}. These approaches rely on refinement of the numerical mesh (or grid) as a approach to quantify the numerical uncertainty in the CFD simulation. 

The present study reviews different UQ strategies for quantifying the aleatory and epistemic uncertainties with a focus on the model-form uncertainties inherent in the eddy-viscosity RANS models. 


The manuscript is laid out as follows. In Section 1, we give an introduction to the sources of uncertainties in CFD simulations and the methods to quantify these uncertainties. Section 2 focuses on the model form uncertainties in turbulence models specifically Reynolds Averaged Navier Stokes models. Section 3 deals with Uncertainty Quantification using forward models. Section 4 deals with Uncertainty Quantification using backward models. Sections 5, 6 and 7 focus on the Eigenspace Perturbation Method (EPM) due to its over arching importance in the UQ community. Section 5 deals with the Eigenvalue perturbation, Section 6 with the Eigenvector perturbation and Section 7 with the perturbations to the turbulent kinetic energy. Section 8 summarizes the manuscript and recommends future directions of research.

\begin{figure} 
\centerline{\includegraphics[width=6.5in]{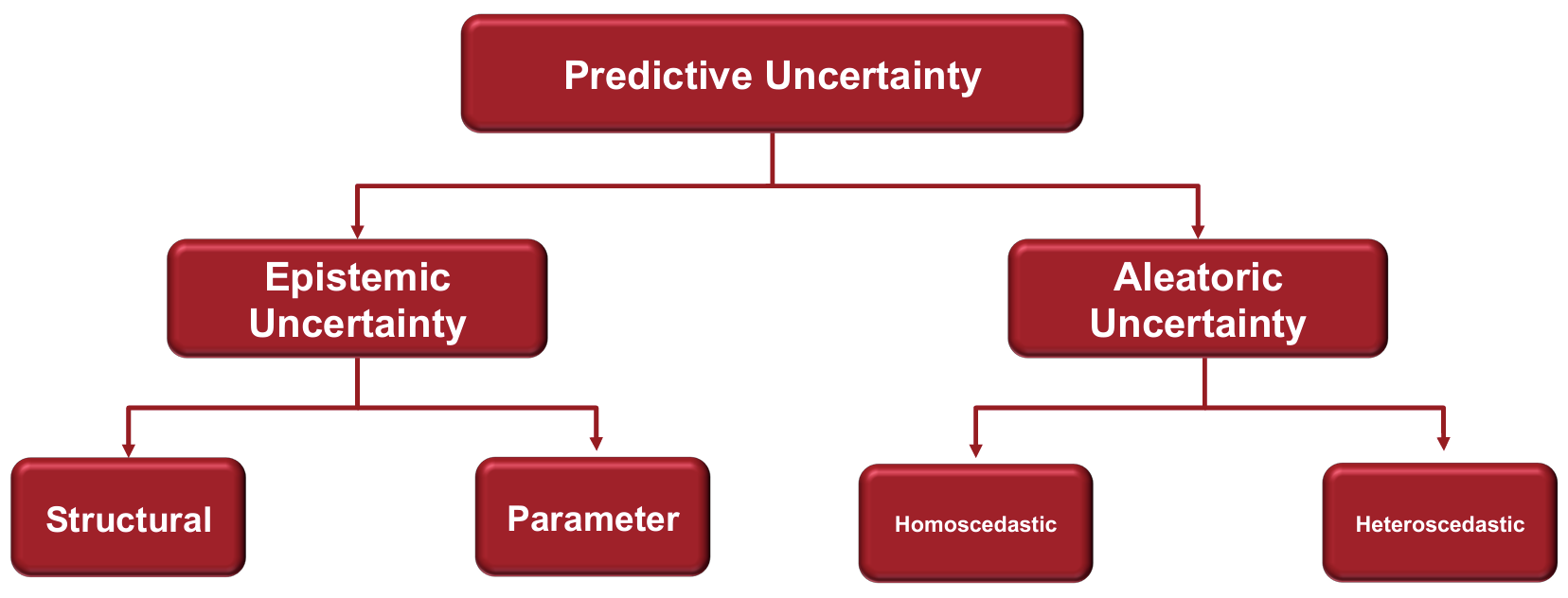}}
\caption{Schematic delineating the different components of a turbulence model's Predictive uncertainty. Different sources of uncertainty can be dominant for different flow problems and even different regions of the same flow problem.}
\label{fig:UQtypes.png}
\end{figure}

\section{Quantifying epistemic uncertainties in RANS models}
In terms of types, predictive uncertainty can be either aleatory or epistemic, as shown in Fig. \ref{fig:UQtypes.png}. Epistemic uncertainty arises due to knowledge missing for describing a quantity, which can be reduced when more knowledge of the physics of flow becomes available. The epistemic uncertainty can be further classified into parametric and non-parametric uncertainties (structural) depending on where uncertainties are introduced. Parametric uncertainties are introduced to the closure model coefficients, and hence less physical insights. In contrast, non-parametric uncertainties are directly intrinsic to the modeled terms that have strong physical insights, such as the eddy viscosity \cite{wang2010quantification}, the source terms of transport equations \cite{singh2016using}, or the Reynolds stress \cite{xiao2016quantifying,ling2016reynolds}. 

From the probability perspective, parametric and non-parametric uncertain quantities of interest can be represented as random variables. It is possible to distinguish the random variables depending on how the random variables are indexed. For instance,  all possible values of a scalar random variable ($X$) can be represented as a vector of random variables $\boldsymbol{X}=\left[X_1, \cdots, X_n\right]$, indexed by integers. Further, a random field $X(y)$ refers to a field of random variables indexed by the spatial coordinate $y$. In addition, it can be referred to as a field of stochastic variables indexed by time coordinate $t$.  

Depending on the uncertainty characteristics, forward (data-free) and backward (data-driven) methods are used to quantify and reduce uncertainty. The forward methods need pre-specified probability distributions on the parametric and non-parametric uncertainties. These uncertainties are then propagated through the governing RANS equations. On the other hand, backward methods assimilate the given observed data to infer the parametric/non-parametric uncertainties. It should be noted that the inferred probability distributions will then be used in the subsequent prediction step as the forward method does.  

On the other hand, aleatory uncertainty results from variability. Aleatory uncertainty is homoscedastic if the variability of a variable has the same finite variance, while the variance is unequal for a heteroscedastic variable, as shown in Fig. \ref{fig:UQtypes.png}. From the frequentist probability perspective, aleatory uncertainties are inherent in stochastic quantities whose variability may be described by probability density function (PDF). While PDF cannot be constructed for quantities with epistemic uncertainty from a frequentist probability perspective. This difficulty does not exist from the Bayesian perspective. In the Bayesian framework, a PDF is constructed depending on the degree of belief as long as sufficient prior information is available. 

\section{Uncertainty quantification through forward method}
Forward methods require a known (prior) probability distribution $p(\theta)$ of uncertain quantities of interest, such as model parameters. As the prior probability distribution is well specified, you can easily sample from high probability regions \cite{ghanem2003stochastic}. Forward methods are good to address uncertainties in model parameters, i.e., propagating the model uncertainties from model parameters to model outputs. Forward methods can be classified into spectral methods \cite{ghanem2003stochastic} and Monte Carlo methods \cite{glasserman2004monte}. Spectral methods use orthogonal basis function to discretize the uncertain space of random variables. For instance, Polynomial Chaos Expansion (PCE) is a valuable tool for constructing an approximate relationship between input parameters and the output of a model in the presence of uncertainty. On the other hand, Monte Carlo methods estimate the uncertainty of the output using input random variables. The big drawback of Monte Carlo methods is its slow convergence rate. The convergence rate is proportional to the number of samples at a rate of $\mathcal{O}\left(N^{-1 / 2}\right)$ \cite{glasserman2004monte}.

\section{Uncertainty quantification through backward method}
Bayesian frameworks estimate uncertainties based on available data using the Bayes' theorem:

\begin{equation}\label{Eq:Bayes}
    p(\theta \mid Z)=\frac{p(Z \mid \theta) p(\theta)}{p(Z)},
\end{equation}

where $p(\theta)$ is the prior, $p(Z \mid \theta)$ is the likelihood, $p(\theta \mid Z)$ is the posterior probability, and the $p(z)$ is total probability of the observed data for normalization. Equation \ref{Eq:Bayes} states that the posterior probability is proportional to the $p(\theta)$ and the $p(Z \mid \theta)$. 

\subsection{Bayesian framework based on Markov Chain Monte Carlo method}
Unlike forward methods that are straightforward using plain Monte Carlo sampling, backward methods based on Bayes' theorem sample from the unknown posterior probability distribution. Since the plain Monte Carlo sampling has difficulty in locating the high probability regions from posterior, the plain Monte Carlo is usually not used with Bayesian frameworks. Alternatively, Markov chain Monte Carlo (MCMC) methods are frequently employed. MCMC methods belong to a category of sequential sampling techniques where the subsequent sampled state relies solely on the current state. This approach enables sampling to concentrate on areas of high probability, occasionally exploring regions of low probability (tails). By using a target distribution, the MCMC algorithm generates samples from that distribution by creating a Markov chain with a stationary distribution that matches the desired target distribution.


\begin{figure} 
\centerline{\includegraphics[width=5.5in]{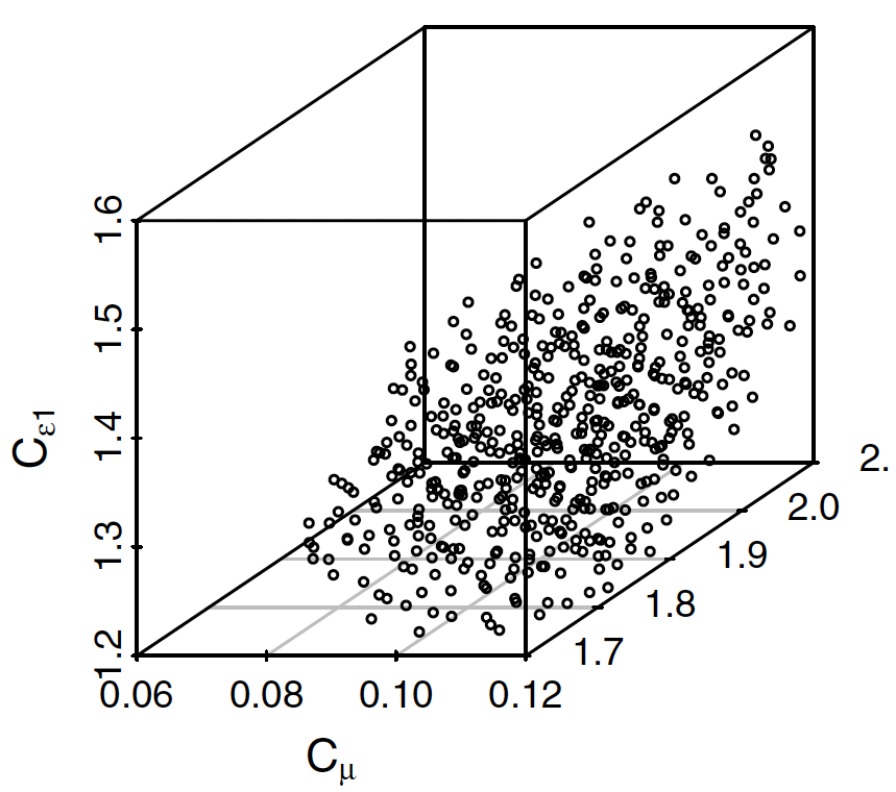}}
\caption{Scatterplot of the set of $525$ points (runs) in the $(C_{\mu}, C_{\epsilon 2}, C_{\epsilon 1})$ space that constitute $\mathcal{R}$ for root-mean-square error (RMSE) liying below the 20th percentile.}
\label{fig:MCMC subspace}
\end{figure}

Although the MCMC stands as the pinnacle of Bayesian inference and posterior sampling, its practical application highly demands a substantial volume of samples for achieving statistical convergence. Typically, the requisite number of samples falls within the range of $\mathcal{O}(10^{5})$ to $\mathcal{O}(10^{6})$, varying based on the shape of the posterior distribution and the efficiency of the sampling process. In the realm of Computational Fluid Dynamics (CFD), each assessment involves simulations lasting hours or even weeks, contingent upon the complexity of the flow configuration.

we can conclude that conducting RANS simulations for every likelihood evaluation within the MCMC sampling is unfeasible. This impracticality doesn't solely stem from the high volume of samples needed but also from the sequential process inherent in traditional MCMC algorithms. In these algorithms, the proposal of the subsequent sample relies on the assessment of the current state's posterior, further complicating the feasibility of conducting full simulations at each step. To solve this difficulty surrogate models are in widespread application in MCMC-based model uncertainty quantification. They serve the purpose of mitigating the substantial computational expenses associated with RANS simulations when evaluating likelihood.

A Bayesian calibration approach \cite{ray2016bayesian,ray2018learning} is adopted to calibrate a RANS model and estimate its model parameters, $\mathbf{C}=(C_{\mu}, C_{\epsilon 2}, C_{\epsilon 1})$. The Bayesian approach is solved using an MCMC method. The joint probability density function of the parameters and the model-data misfit, denoted as $P\left(\boldsymbol{C}, \sigma^2 \mid \boldsymbol{y}_e\right)$, is conditional on the observed data $y_{e}$. The prior beliefs about the distribution of $\boldsymbol{C}$ and $\sigma^{2}$ are represented by $\Pi_{1}(\boldsymbol{C})$ and $\Pi_{2}(\sigma^{2})$ respectively. The likelihood of observing $y_{e}$, given a parameter setting $\boldsymbol{C}$, denoted as $\mathcal{L}(y_e \mid \boldsymbol{C})$, is expressed as follows:

\begin{equation}\label{Eq:likelihood}
    \mathcal{L}\left(\boldsymbol{y}_e \mid \boldsymbol{C}, \sigma^2\right) \propto \frac{1}{\sigma^{N_p}} \exp \left(-\frac{\left\|\boldsymbol{y}_e-\boldsymbol{y}_m(\boldsymbol{C})\right\|_2^2}{2 \sigma^2}\right)
\end{equation}

Based on the Bayes' theorem in Eqn. \ref{Eq:Bayes}, the updated (calibrated) posterior distribution of $(\boldsymbol{C}, \sigma^{2})$ can be defined as:

\begin{equation}\label{Eq:posterior}
    \begin{aligned} P\left(\boldsymbol{C}, \sigma^2 \mid \boldsymbol{y}_e\right) & \propto \mathcal{L}\left(\boldsymbol{y}_e \mid \boldsymbol{C}, \sigma^2\right) \Pi_1(\boldsymbol{C}) \Pi_2\left(\sigma^2\right) \\ & \propto \frac{1}{\sigma^{N_p}} \exp \left(-\frac{\left\|\boldsymbol{y}_e-\boldsymbol{y}_m(\boldsymbol{C})\right\|_2^2}{2 \sigma^2}\right) \Pi_1(\boldsymbol{C}) \Pi_2\left(\sigma^2\right).\end{aligned}
\end{equation}

The MCMC method can be used to draw samples ($O(10^{4})$) of $\{\boldsymbol{C}, \sigma^{2} \} $ to reconstruct the posterior probability distribution $P\left(\boldsymbol{C}, \sigma^2 \mid \boldsymbol{y}_e\right)$. Histograms or Kernel density estimation can facilitate the visualization of $P\left(\boldsymbol{C}, \sigma^2 \mid \boldsymbol{y}_e\right)$ \cite{silverman2018density}. When the MCMC chain is converged to a stationary posterior probability distribution, a sufficient number of samples of $\{\boldsymbol{C}, \sigma^{2} \} $ is required, and an algorithm is needed to determine the sufficiency, e.g., \cite{raftery1996implementing}. 

Due to the high volume of samples and hence extremely high computational cost associated with the MCMC method, i.e., each of the $O(10^{4})$ samples requires a RANS simulation to provide $\mathbf{y_{m}(C)}$ in Eqn. \ref{Eq:posterior} and hence impractical. As a result, a RANS model can be replaced with a polynomial surrogate to significantly reduce the computational cost via mapping the dependence of the uncertain quantities of interest on $\mathbf{C}$. The (polynomial) surrogate model is constructed involving $\boldsymbol{C}$, i.e., $\boldsymbol{C \in R}$. Note that the surrogate method is limited in its applicability to state spaces characterized by lower dimensions. This mapping remains accurate within an acceptable margin of error in the support of $\Pi_1(\boldsymbol{C})$.  Usually, we can establish the bounds of the uncertain parameter space $\mathbf{C}$; however, parameter combinations chosen randomly from this space might be physically unrealistic. It could potentially lead to a crash in RANS simulations. Therefore, we need the selection of a part $\mathbf{R}$ of $\mathbf{C}$ space that contains the values leading to physically realistic flowfields. Consequently, the development of an informative prior is required to restrict the values of parameters to a Region $\mathcal{R}$. Figure \ref{fig:MCMC subspace} shows the subset $\boldsymbol{R}$ space is identified to exclude a large portion of $\boldsymbol{C}$ \cite{ray2016bayesian}. The subset $\boldsymbol{R}$ space is identified through the evaluation of the root-mean-square error (RMSE) between the vorticities produced by each of the simulations and the experimental counterpart on the crossplane \cite{ray2016bayesian}. Among the selected number of runs, the uncertain values of $(C_{\mu}, C_{\epsilon 2}, C_{\epsilon 1})$ that have led to the RMSE lying below the $20th$ percentile identify $\boldsymbol{R}$, as shown in Fig. \ref{fig:MCMC subspace}.






\subsection{Approximate Bayesian framework based on Ensemble Kalman filter method}
The MCMC approach gives a highly precise sampling of the posterior distribution; however, it requires a substantial number of samples. In cases where exact probability isn't crucial and only lower-order statistical moments like the mean and variance hold significance, alternative approximate Bayesian inference techniques come into play. These methods rely on the maximum a posteriori (MAP) probability estimate to identify the mode (peak) of the posterior distribution, rather than capturing the entire distribution itself. In this review, our attention is directed toward a specific MAP method known as the ensemble Kalman filter method, which has been widely used in Bayesian frameworks \cite{iglesias2013ensemble,xiao2016quantifying}. The original Ensemble Kalman Filter (EnKF) is a sequential filter method, from which the model is integrated forward in time. Whenever the observed data are available the predictions are compared with the observed data to reinitialize the model before the integration continues. The original EnKF method has been widely applied to various applications of data assimilation for state and parameter estimation, especially in oceanography \cite{evensen1996assimilation}, reservoir modeling \cite{aanonsen2009ensemble} and weather forecasting \cite{houtekamer2001sequential}. 

Inspired by the original EnKF, a recent iterative EnKF (IEnKF) involves multiple iterations of the EnKF update step. After the initial EnKF update, IEnKF performs additional iterations, where each iteration refines the estimation by repeatedly assimilating observations and updating the ensemble. This iterative process helps to improve the accuracy of the state estimation, especially in situations where the original EnKF might exhibit limitations or inaccuracies. The system state is initially expanded to encompass both the observable and physical states $x(t)$, e.g., velocity, pressure, and/or turbulence kinetic energy fields, and parameters $\mathbf{\theta}$, e.g., model coefficients or viscosity field, which remain unobservable and require inference. This type of system state is referred to as an augmented state. Ensemble Kalman-based methods essentially update an ensemble of an augmented state via the Kalman formulation which combines the model prediction and observed data at a given time \cite{evensen2003ensemble,evensen2009data}. 

To infer the unknown of $u$ using an IEnKF, we need the observed data expressed as

\begin{equation} \label{Eq:Observations}
    y = \mathcal{G}(u) + \epsilon,
\end{equation}

where $\mathcal{G}$ is the forward response operator that maps $u$ to the observations space. Note that $\mathcal{G}$ gives a forward response from $u$ that is computed by a PDE system describing a physical system. $\epsilon$ is assumed a standard normal distribution with zero mean and known covariance $Cov$. To solve the inverse problem, let $X$ and $Y$ be the prediction space and the observation space, respectively. The artificial dynamics of an augmented state can be constructed as

\begin{equation} \label{Eq:AugmentStateMapping}
    \Xi\left(z\right)=\begin{pmatrix} y\\ \mathcal{G}(u). \end{pmatrix} \quad \text{for} \quad z = \begin{pmatrix}
        u\\ p 
    \end{pmatrix} \in Z.
\end{equation}

In Eqn. \ref{Eq:AugmentStateMapping}, $z$ is the augmented state being mapped to $Z$ space. Therefore, the artificial dynamics can be expressed as 

\begin{equation}\label{Eq:airtificaldynamics_AugState}
    z_{n+1} = \Xi\left( z_{n} \right).
\end{equation}

The observed data in Eqn. \ref{Eq:Observations} can be assumed based on the artificial dynamics to be

\begin{equation} \label{Eq:y_n+1}
    y_{n+1} = Hz_{n+1} + \epsilon_{n+1},
\end{equation}

where $H = (0, I)$ is the measurement function or projection operator mapping $Z$ space to $Y$ space. Then the objective of the EnKF approach is to estimate the augmented state in Eqn. \ref{Eq:airtificaldynamics_AugState}, from which the unknown $u$ in Eqn. \ref{Eq:Observations} can be computed. At each iteration, an ensemble of particles is updated by combining the augmented state (Eqn. \ref{Eq:airtificaldynamics_AugState}) with the observed data (Eqn. \ref{Eq:Observations}). The iteration index $n$ in Eqns. \ref{Eq:airtificaldynamics_AugState} and \ref{Eq:y_n+1} represent an artificial time, while the real time pertains to $\mathcal{G}$ and is not related to $n$. 

From Eqns. \ref{Eq:Observations} to \ref{Eq:y_n+1}, the inverse problems are ill-posed, hence regularization is required by incorporating prior knowledge of $u$ in a finite-dimensional subspace $\mathcal{A}$, i.e., $\mathcal{A} \in X$. The EnKF-based solution in Eqn. \ref{Eq:Observations} remains in $\mathcal{A}$. EnFK framework constructs an interacting ensemble of particles $\{z^{(j)}_{n}\}^{J}_{j=1}$, which is used to estimate the unknown $u$ as follows:

\begin{equation}\label{Eq:unknownu}
    u_{n} \equiv \frac{1}{J}  \sum_{j=1}^{J} u^{(j+1)}_{n} = \frac{1}{J}  \sum_{j=1}^{J} H^{\perp}z^{(j)}_{n}.
\end{equation}

The EnKF method requires a first guess of $\{z^{(j)}_{0}\}^{J}_{j=1}$ to be initiated. The ensemble $\{z^{(j)}_{0}\}^{J}_{j=1}$ can be created by forming an ensemble $\{\psi^{(j)}\}^{J}_{j=1}$ within the $\mathcal{A}$ space where the solution of the unknown $u$ is sought. Then, we can set

\begin{equation} \label{Eq:z_{0}}
    z^{(j)}_{0} = \begin{pmatrix} \psi^{(j)} \\ \mathcal{G}(\psi^{(j)}). \end{pmatrix}.
\end{equation}

The corresponding $u_{0}$ (see Eqn. \ref{Eq:unknownu}) is just the mean of the initial ensemble in $\mathcal{A}$ space. The initial ensemble $\{\psi^{(j)}\}^{J}_{j=1}$ can be constructed from the available prior knowledge, e.g., Gaussian $N(\overline{u}, P)$ and $\overline{u} + \sqrt{\lambda_{j}}\phi_{j}$, where $(\lambda_{j},\phi_{j})$ represent eigenvalue and eigenvector Paris of $P$ with eigenvalue in descending order, that is, the Karhunen-Alfred Lo\'eve (KL) basis.

With $\{z^{(j)}_{0}\}^{J}_{j=1}$ being specified, the EnKF method will update the ensemble of particles $\{z^{(j)}_{n}\}^{J}_{j=1}$ iteratively. Overall, the iterative EnKF consists of two steps: the prediction step and the update step at each iteration. The prediction step propagates the ensemble of particles based on Eqn. \ref{Eq:airtificaldynamics_AugState}. As the prediction step uses a forward model, the ensemble being mapped to the observed data space brings the information of the forward model.  The prediction step is defined as:

\begin{equation} \label{Eq:ensemble_AugState}
    \widehat{z}^{(j)}_{n+1} = \Xi\left( z^{j}_{n} \right).
\end{equation}

From Eqn. \ref{Eq:ensemble_AugState}, the augmented state $\widehat{z}^{(j)}_{n+1}$ can be defined as its mean and covariance as follows:

\begin{equation}
    \hat{z}_{n+1} = \frac{1}{J}  \sum_{j=1}^{J} \widehat{z}^{(j)}_{n+1} 
\end{equation}

\begin{equation}
    P_{n+1} = \frac{1}{J} \sum_{j=1}^{J} \widehat{z}^{(j)}_{n+1}(\widehat{z}^{(j)}_{n+1})^{T} - \overline{z}_{n+1}\overline{z}^{T}_{n+1}.
\end{equation}

In the update step, the calculation of the Kalman gain blending covariance matrices from the prediction and the observed data is defined as follows: 

\begin{equation}
    K_{n+1} = P_{n+1}H^{\star}(HP_{n+1}H^{\star}+R)^{-1},
\end{equation}

where $H^{\star}$ is the adjoint operator of $H$. With Kalman gain, the updated augmented state for each ensemble member can be computed as follows:

\begin{equation}
    z_{n+1}^{(j)} = I\widehat{z}^{(j)}_{n+1} + K_{n+1}(y^{(j)}_{n+1} - H\widehat{z}^{(j)}_{n+1})
\end{equation}

where $y^{(j)}_{n+1} - H\widehat{z}^{(j)}_{n+1}$ is the residual, and 

\begin{equation}
    y^{(j)}_{n+1} = y + \epsilon^{(j)}_{n+1}. 
\end{equation}

In the observed data space, the update step compares the mapped ensemble with the noisy observed data; it attempts to correct the ensemble to match the observed data better. From Eqn. \ref{Eq:unknownu}, the mean of the unknown $u$ is updated as follows:

\begin{equation}\label{Eq:unknownu_update}
    u_{n+1} \equiv \frac{1}{J}  \sum_{j=1}^{J} u^{(j+1)}_{n+1} = \frac{1}{J}  \sum_{j=1}^{J} H^{\perp}z^{(j)}_{n+1}.
\end{equation}

The stopping criterion for the EnKF method is based on the discrepancy principle, from which the EnKF method is terminated for the first $n$ such that 

\begin{equation}\label{Eq:convergence}
    \left\Vert  y- \mathcal{G}(u_{n}) \right\Vert_{R} \leq \tau \left\Vert \epsilon^{\star} \right\Vert_{R} \quad\quad \text{for some $\tau > 1$},
\end{equation}

where $\epsilon^{\star}$ is the noise in the true observed data.

\subsection{Physics-informed IEnKF in Bayesian framework}
\begin{figure} 
\centerline{\includegraphics[width=5.5in]{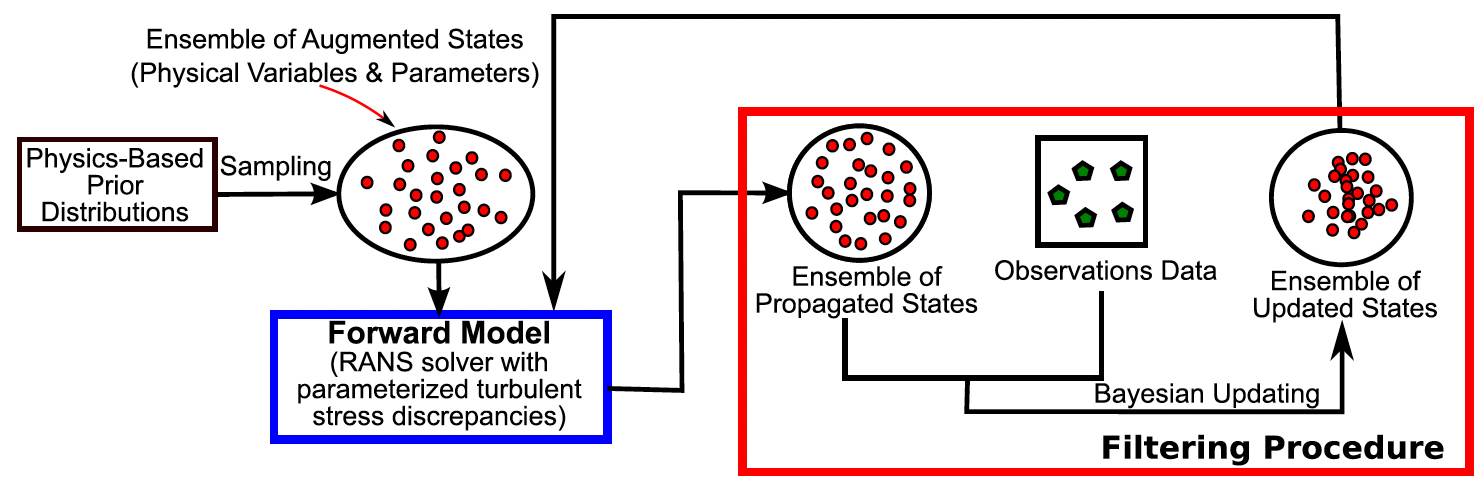}}
\caption{IEnKF algorithm in an approximate Bayesian inference framework.}
\label{fig:XiaoBayesianFramework.png}
\end{figure}

IEnKF method incorporates both physics-based prior knowledge and observed data within a Bayesian framework. This process aims to update the posterior distributions of an augmented state. This iterative scheme is shown in Fig. \ref{fig:XiaoBayesianFramework.png}. In this particular example, the augmented state is $\mathbf{X} = [\mathbf{u}, \tau^{param}]^{T}$ where $\mathbf{u}$ represents the velocity field, $\tau$ represents the Reynolds stress field, and $param$ means parameterized. The goal is to infer/reconstruct Reynolds stresses by comparing the baseline prediction for velocity fields with the available observed data. The physics-based priors introduce physical meaning to $\tau^{param}$, and the initial prior ensemble of augmented states ${\mathbf{u}}^{N}_{j=1}$ and ${\mathbf{\tau^{param}}}^{N}_{j=1}$ are generated from sampling the baseline RANS prediction \cite{xiao2016quantifying}, where $\mathbf{N}$ is the sample size. Note that this baseline simulation is performed only once for initialization. Forward model includes an ensemble of ${\mathbf{\tau^{param}}}^{N}_{j=1}$. For each sample in the ensemble ${\mathbf{\tau^{param}}}^{N}_{j=1}$, the corresponding ${\mathbf{u}}^{N}_{j=1}$ is obtained by solving the RANS equations. Kalman filtering procedure (Kalman update step) compares ensemble mean of ${\mathbf{u}}^{N}_{j=1}$ with available observed data of velocity. As a result, the ensemble augmented states ${\mathbf{x}}^{N}_{j=1}$ is updated. A stopping criterion is based on the Eqn. \ref{Eq:convergence} when statistical convergence of the ensemble is achieved.

In the example shown in Fig. \ref{fig:XiaoBayesianFramework.png}, RANS equations (PDEs) are considered exact; however, they include unknown Reynolds stress fields, which are not solved directly and must be approximated using a turbulence model. The Reynolds stress field is considered the dominant source of uncertainty in RANS modeling \cite{oliver2009uncertainty}. Therefore, these uncertainty Reynolds stress fields can be referred to as latent physical fields \cite{strofer2020enforcing}. Latent physical fields also include initial conditions of the fields of interest, constant physical properties such as density field or viscosity field, and any physical fields related to the fields of interest. Latent physical fields can be inferred using observed data of the field of interest. 

\begin{equation}
    \mathcal{M}(u; l) = 0
\end{equation}

where $\mathcal{M}$ represents the governing PDEs of a dynamic system, $u$ represents fields of interest, and $l$ represents latent fields. 

The fields of interest can be related to the latent fields as

\begin{equation}
    u = \mathcal{F}(l).
\end{equation}

As the PDEs are considered exact, latent fields are the only source of uncertainty. Therefore, the predicted fields of interest can be improved through inferring the latent uncertain fields. Many studies have been conducted to infer the latent uncertain fields using sparse observations \cite{kennedy2001bayesian, oliver2015validating,duraisamy2019turbulence,xiao2019quantification}. The selected statistical model describing the prior distribution, especially its covariance kernel, can significantly influence Bayesian inference. In the example shown in Fig. \ref{fig:XiaoBayesianFramework.png} \cite{xiao2016quantifying}, the prior for the latent Reynolds stress field is modeled as Gaussian process: $\mathcal{G}\mathcal{P}(0, k)$, where

\begin{equation}
    K(x,x^{'}) = \sigma(x)\sigma(x^{'})exp(-\frac{|x-x^{'}|^{2}}{L^{2}})
\end{equation}

is the covariance kernel for two locations $x$ and $x^{'}$. The variance $\sigma(x)$ varies spatially to reflect large discrepancies in certain areas in the domain. The $L$ is the correction length based on the local turbulence length scale. Bayesian inference problems are ill-posed, Gaussian process might not be able to infer the true latent field, although updated fields of interest might be improved. Therefore, more physically-realistic global constraints can be enforced via a more complex statistical model than a simple Gaussian process. The random latent field's representation is selected in a way that ensures automatic adherence to the physical constraints in any realization of the field. For instance, enforcing a positivity constraint could involve modeling the latent field as a lognormal process.

   

\begin{subequations}
\begin{equation}
   \tau = e^{\delta}
\end{equation}    
\begin{equation}
   \delta \sim \mathcal{G}P(log{(\Tilde{\tau}),K}),
\end{equation}
\end{subequations}

where $\Tilde{\tau}$ is the baseline solution. More constraints, particularly regarding covariance kernel can be found in \cite{strofer2020enforcing}.




\subsection{Forward propagation of uncertainty}
In this section the discussion focuses on the uncertainties involved with the prediction tasks as opposed to inference uncertainties involved in the backward propagation. The prediction uncertainties include epistemic and aleatory uncertainties. The epistemic uncertainties are because of the structure of the turbulence model used and its shortcomings in expressing turbulence physics. The aleatory uncertainties are because of measurement errors in initial or boundary conditions. 

\subsubsection{Model-form/structural uncertainty}
The model form uncertainty in turbulence models is the most important source of uncertainty in CFD simulations of turbulent flows. This has been called the greatest challenge in the design of Aerospace vehicles \cite{zang2002needs}. Within RANS models, the approximation of Reynolds stress terms relies on the Boussinesq turbulent-viscosity hypothesis, where anisotropic Reynolds stresses are linearly related to the mean rate of strain. These models, also known as linear eddy viscosity models, have limitations well-documented in handling intricate flow scenarios, such as those with pronounced streamline curvature, separation, and reattachment. While Large Eddy Simulations (LES) or Direct Numerical Simulations (DNS) offer highly accurate solutions for such complexities, their computational demands in terms of time and cost are often prohibitive, especially for high-Reynolds number flows. Consequently, assessing the uncertainties within the RANS model emerges as a valuable alternative for refining predictive capabilities in engineering applications. The consideration of more expensive LES or DNS simulations is warranted only if the uncertainties inherent in the RANS model exceed acceptable limits.

RANS modeling continues to remain popular and widespread across various engineering fields due to its rather low computational cost and acceptable robustness. As a comparison RANS models involve the solution of two-equations for two unknowns. The next level of Reynolds Stress Modeling involves the solution of eight equations for eight unknowns. The accuracy of Reynolds Stress Models is also limited at best \cite{pope2001turbulent}. Approaches like DNS and LES invoke orders of magnitude higher computational costs. In this setting RANS models are popular due to their compromise between high robustness and low computational cost. This lowering of the computational cost is achieved using approximations and simplifications in the developing of the RANS models leading to severe model-form uncertainties in the predictions of RANS models. Different researchers have used different approaches to quantify the model form uncertainty from turbulence models. For example instead of using a single turbulence model researchers have used an ensemble of turbulence models and used the variability in their predictions as an estimate of the turbulence model uncertainty \cite{vuruskan2019impact, vuruskan2022impact}. Other researchers have tried using more advanced turbulence models like Reynolds Stress Models (RSM) \cite{pope2001turbulent}. Besides adding to the computational cost this does not remove turbulence model uncertainty because all turbulence models suffer from model form uncertainties \cite{yuan2019single}. Iaccarino and co-workers \cite{emory2011modeling,iaccarino2017eigenspace,gorle2013framework,emory2013modeling,gorle2014deviation, Emory2014Thesis} developed an approach to estimate the model-form uncertainty in RANS modeling by decomposing the Reynolds stress field and perturbing its magnitude, eigenvalues, and eigenvectors toward their limiting states within the physically realizable range. This framework focuses only on the forward propagation of uncertainties in the Reynolds stress field.

\begin{figure} 
\centerline{\includegraphics[width=5.5in]{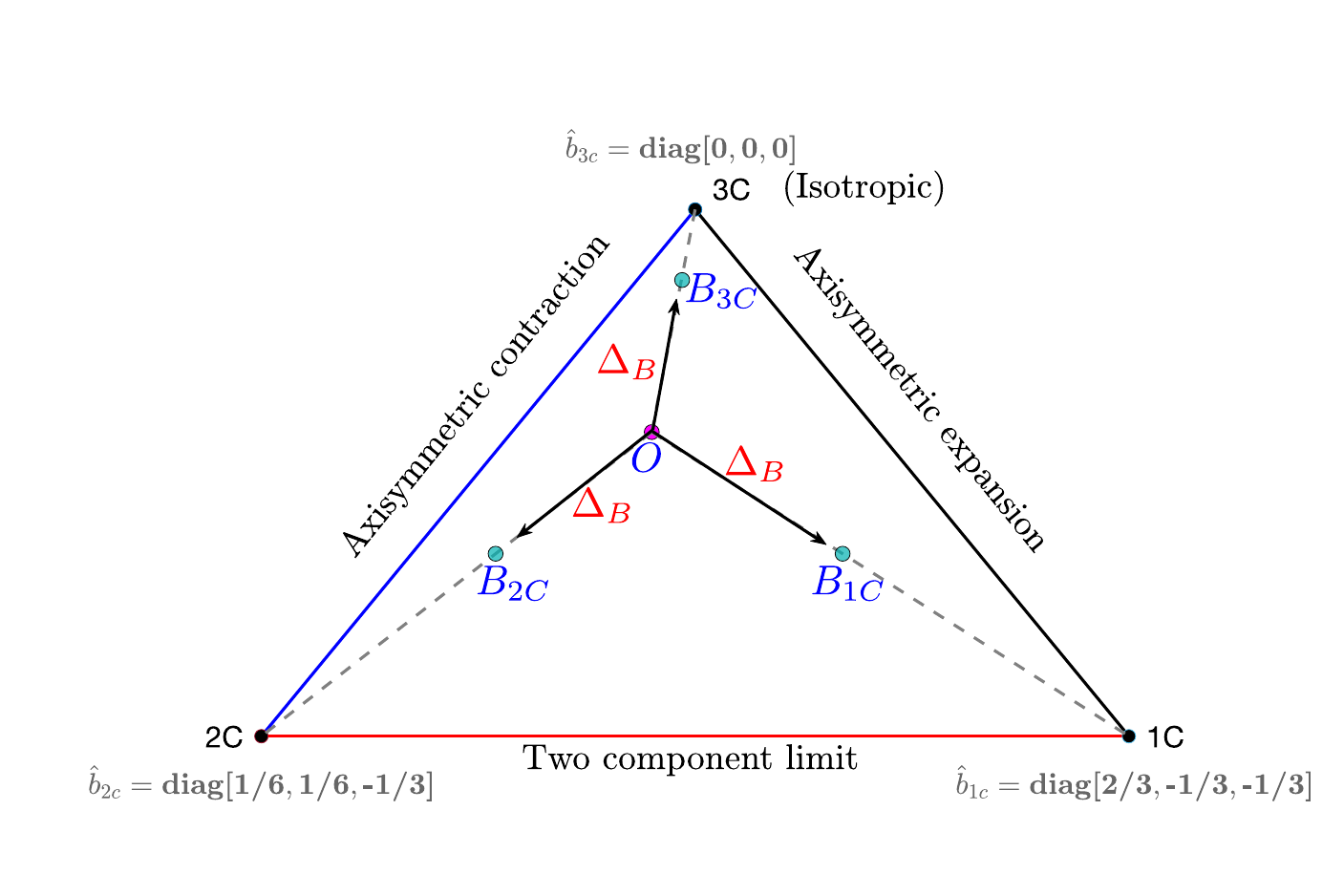}}
\caption{Schematic of the Barycentric Triangle. All realizable states of the Reynolds stresses lie inside or on the triangle.}
\label{fig:BMap_Sketch.pdf}
\end{figure}

RANS modeling aims at closing the time-averaged momentum equations by approximating the unknown Reynolds stresses, which gives deterministic predictions for flow behaviors under specific conditions. The Reynolds stresses $R_{ij}=\langle u_iu_j \rangle$ are the important prediction for turbulence models. Reynolds stress decomposed into the anisotropic and isotropic components with

\begin{equation}
R_{ij}=2k(b_{ij}+\frac{\delta_{ij}}{3}),
\end{equation}
where $k(=\frac{R_{ii}}{2})$ is the turbulence kinetic energy, $b_{ij}(=\frac{R_{ij}}{2k}-\frac{\delta_{ij}}{3})$ is the anisotropy tensor. 

An alternative way to close the time-averaged equations is to replace the unknown stresses by introducing the uncertainty intervals. As such, instead of yielding explicit estimates, possible behaviors are estimated. Based on the concept of Reynolds stress tensor satisfying physical realizability constraints \cite{schumann1977realizability, lumley1979computational, durbin1994realizability, banerjee2007presentation}, Banerjee \textit{et al.} 
introduced the concept of a two-dimensional barycentric triangle to facilitate visual representation of the realizable Reynolds stress states. Reynolds stress states must be constrained within this Barycentric map \cite{banerjee2007presentation}. In spectral space the Reynolds stress anisotropy is represented via
\begin{equation}
b_{in}v_{nl}=v_{in}\Lambda_{nl},
\end{equation}
where $v_{nl}$ is a matrix of orthonormal eigenvectors, $\Lambda_{nl}$ is the diagonal matrix of eigenvalues $\lambda_{k}$ and is traceless. Multiplication by $v_{jl}$ gives $b_{ij}=v_{in}\Lambda_{nl}v_{jl}$. This is substituted into the spectral form of the Reynolds stress anisotropy to give
\begin{equation}
R_{ij}=2k(v_{in}\Lambda_{nl}v_{jl}+\frac{\delta_{ij}}{3}).
\end{equation}

The $v$ and $\Lambda$ are ordered so $\lambda_{1}\geq\lambda_{2}\geq\lambda_{3}$. Using this the shape, orientation and amplitude of the Reynolds stress ellipsoid are reflected by the turbulence anisotropy eigenvalues $\lambda_l$,  eigenvectors $v_{ij}$ and the turbulent kinetic energy $k$.

The perturbed Reynolds stress tensor can be expressed as
\begin{equation}
R_{ij}^*=2k^* (\frac{\delta_{ij}}{3}+v^*_{in}\Lambda^*_{nl}v^*_{lj})
\end{equation}
where $^*$ represents quantities after perturbation. $k^*=k+\Delta k$ is the perturbed turbulent kinetic energy, $v^*_{in}$ is the perturbed eigenvector matrix, and, $\Lambda^*_{nl}$ is the diagonal matrix of perturbed eigenvalues, $\lambda_l^*$. 

 For eigenvalue perturbation, barycentric map \cite{banerjee2007presentation} is used to enforce the realizability constraints on $\left\langle u_{i}u_{j} \right\rangle$, as shown in Fig. \ref{fig:BMap_Sketch.pdf}. This approach was proposed by Pecnik and Iaccarino \cite{emory2011modeling}. On the three corners of the map, $1c$, $2c$ and $3c$ are the three extreme states of componentiality of $\left\langle u_{i}u_{j} \right\rangle$. Physically, $1c$ represents a ``rod-like'' principal fluctuation in one direction, $2c$ represents a ``pancake-like'' principal fluctuations of the same intensity in two directions, and $3c$ represents a ``spherical'' principal fluctuations of the same intensity in three directions. Given an arbitrary point $\mathbf{x}$ within the barycentric map, any realizable $\left\langle u_{i}u_{j} \right\rangle$ can be determined by a convex combination of the three vertices $\mathbf{x}_{i c}$ (limiting states) and $\lambda_{l}$. The perturbed eigenvalues are given via $\lambda_l^*=B^{-1}\mathbf{x^*}$. $\mathbf{x^*}=\mathbf{x} +\Delta_B(\mathbf{x^t}-\mathbf{x})$ is the representation of the perturbation in the barycentric triangle with $\mathbf{x}$ being the unperturbed state in the barycentric map, $\mathbf{x^*}$ representing the perturbed position, $\mathbf{x^t}$ representing the state perturbed toward and $\Delta_B$ is the magnitude of the perturbation. In this context, $\lambda_l^*=B^{-1}\mathbf{x^*}$ can be simplified to $\lambda_l^*=(1-\Delta_B)\lambda_l + \Delta_B B^{-1}\mathbf{x^t}$. Here, $B$ defines a linear map between the perturbation in the barycentric triangle and the eigenvalue perturbations.With the three vertices $x_{1C}$ , $x_{2C}$ , and $x_{3C}$ as the target states, we have $B^{-1}x_{1C} = (2/3, -1/3,-1/3)^T$ , $B^{-1}x_{2C} = (1/6, 1/6,-1/3)^T$ , and $B^{-1}x_{3C} = (0,0,0)^T$ .

\section{Eigenvalue perturbation}
The Eigenvalue perturbation represents the Reynolds stress tensor predicted by the turbulence model via its Spectral decomposition. It then perturbs the eigenvalues of this Spectral decomposition after which the Reynolds stress tensor corresponding to this perturbed Spectral decomposition is reconstituted in physical space. In overall terms the model's predicted Reynolds stress tensor can be represented as an ellipsoid. The Eigenvalue perturbation changes the \textit{shape} of this ellipsoid. 

The need for the Eigenvalue perturbation arises from the inability of eddy viscosity based turbulence models to represent the anisotropy of turbulent flows. On the Barycentric triangle the Reynolds stress predictions of eddy viscosity based turbulence models can be mapped to a single line, referred to as the Plane Strain Line. In real life turbulent flows the Reynolds stresses in a turbulent flow can lie anywhere inside the Barycentric triangle. This limitation of turbulence models means that the models are not able to represent turbulent flows properly. The illustrative flow example is fully developed turbulent pipe flow in pipes with cross-sections that is not circular. These flow show a secondary flow  where the streamwise mean velocity leads to a difference in the normal stresses
in the plane transverse to the streamwise direction. Eddy viscosity based turbulence models are not able to capture this difference in the normal stresses and the secondary flows are absent in the turbulence model predictions. 

\begin{figure} 
\centerline{\includegraphics[width=\textwidth,trim={0 5.5cm 0 5cm},clip]{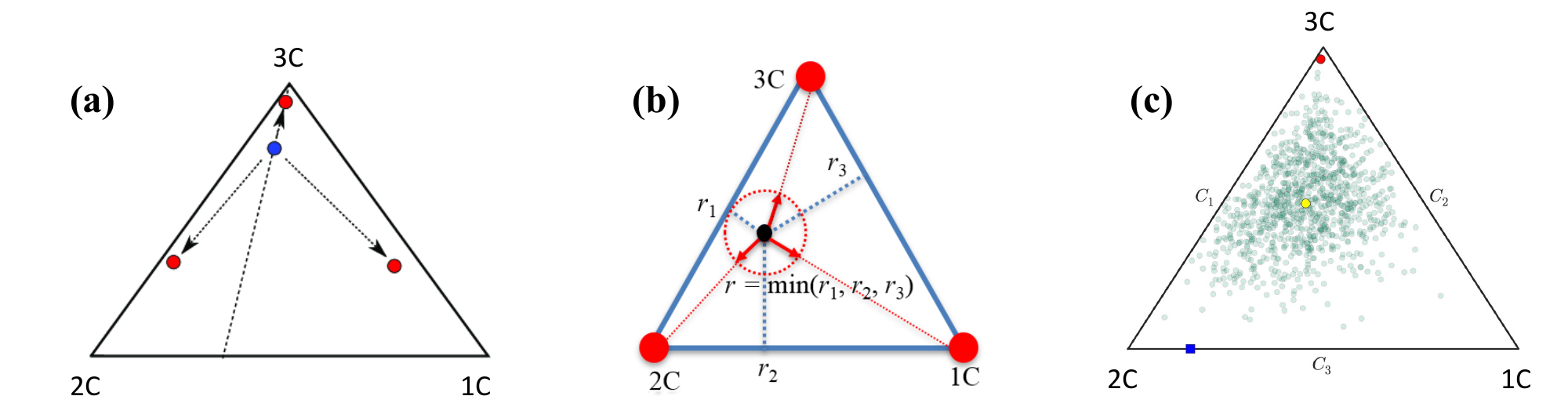}}
\caption{Different approaches used for the eigenvalue perturbation. (a) Uniform relative perturbations towards the limiting states. (b) Non-uniform perturbations towards the limiting states. (c) Random matrix approach with eigenvalue perturbations not toward the limiting states.}
\label{fig:eigenvalueperturbations}
\end{figure}

All the realizable states of turbulence anisotropy lie inside the Barycentric triangle. To maintain realizability the Eigenvalue perturbations are constrained so that the perturbed eigenvalues are mapped to a point inside the Barycentric triangle. This is the only constraint on the eigenvalue perturbation. Theoretically any framework to execute these perturbations is valid as long as it meets this constraint. There are three approaches that have been used:
\begin{enumerate}
    \item Perturbations towards the vetrices of the Barycentric triangle: This is the most commonly used approach for eigenvalue perturbations. Using this approach three perturbed states (towards the 1C, 2C and 3C states of anisotropy) are satisfactory. The magnitude of these perturbations can be either based on domain knowledge of the turbulent flow or predicted by Machine Learning models. 
    These perturbation can be uniform or non-uniform. The uniform case that is most common perturbs towards the 1C, 2C and 3C states by an equal relative magnitude. The non-uniform perturbation towards the vertices of the Barycentric triangle is less common. The central idea for the non-uniform perturbation is that the magnitude of the eigenvalue perturbation should not be the same along all directions for all points in the flow domain. This non-uniform eigenvalue perturbation offers an alternative, where the proximity of the predicted Reynolds stresses in calculated to the sides of the Barycentric triangle to adjudicate the perturbation magnitude \cite{huangnonuniform}. This leads to a variable magnitude of perturbation for each point in the flow domain. 
    
    \item Random perturbations over the Barycentric triangle: In this approach a probability distribution is defined over the Reynolds stress tensor and samples are drawn from this probability density function to create an ensemble \cite{xiao2017random}. This involves a probabilistic model of a random field of positive semi-definite matrices with specified mean and correlation structure, using the maximum entropy principle. To sample from this random matrix distribution, Gaussian probability density functions with the covariance function are generated. These samples from this probability density function are mapped to the space of tensors of appropriate dimension. 
    
    \item Single perturbations towards a corrected state: In the approaches discussed before the magnitude of the perturbation is determined by the user or a data driven approach. The direction is independent of the user or the ML algorithm. The direction of the perturbations can be towards the limiting states of turbulence or in random directions. Some investigators have tried to use machine learning models to predict both the magnitude of the perturbation and the direction of the perturbation. The magnitude of the perturbation shows the absolute distance between the RANS model's predictions and the high fidelity data (LES/DNS) on the Barycentric triangle and the direction shows the alignment of this discrepancy. Given both the direction and the magnitude, this generates a perturbation vector. This perturbation vector can be used to correct the RANS model's predictions to be closer to the eigenvalues in the high fidelity data. 
\end{enumerate}

Various investigators have used data driven Machine Learning models to augment the eigenvalues perturbations \cite{chu2023multi, heyse2021data, heyse2021estimating, matha2023evaluation}. But Machine Learning models have a tendency to overfit. To improve generalization of data driven approaches the community needs to incorporate physics based information into the Machine Learning model. 

In classical physics based models this has been carried out using Realizability conditions \cite{schumann1977realizability, lumley1979computational, du1977realizability}. Realizability requirements are the reason that the eigenvalue perturbations are constrained inside the Barycentric triangle. While this constraint is necessary it may not be sufficient. The constraining of the perturbed eigenvalues inside the Barycentric triangle only constraints the state of the final Reynolds stress tensor. The perturbations define new variants of the underlying RANS models and we need to constrain the Reynolds stress tensor dynamics that are established by these perturbed variants \cite{mishra2014realizability}. Some constraints on the Reynolds stress tensor dynamics due to eigenvalue perturbations have been unearthed in previous investigations. For example \cite{mishra2019theoretical} have shown that just constraining of the perturbed eigenvalues inside the Barycentric triangle is insufficient and can lead to un-physical results as well as numerical instability. 

Another outstanding question is also about realizability. The restriction of the eigenvalue perturbations to stay inside the Barycentric triangle maintains realizability of the predicted Reynolds stresses but does not enforce realizability of the Reynolds stresses. Eddy viscosity based turbulence models predict unrealizable Reynolds stresses very often. These states lie outside the Barycentric triangle and remain unrealizable even after the eigenvalue perturbations. The dynamics of realizable Reynolds stresses have been analysed deeply but not for these predictions where the predictions of the RANS models are unrealizable to start with. Further studies are very needed for this topic. 

\section{Eigenvector perturbation}
The Eigenvector perturbation represents the Reynolds stress tensor predicted by the turbulence model via its Spectral decomposition. Then it perturbs the eigenvectors of this Spectral decomposition, after which the Reynolds stress tensor corresponding to this perturbed Spectral decomposition is reconstituted in physical space. In broad terms, the Reynolds stress tensor can be represented as an ellipsoid. The Eigenvector perturbation changes the \textit{alignment} of this ellipsoid. 

The need for the Eigenvector perturbation arises from the inability of eddy viscosity based models to predict a Reynolds stress tensor with eigenvectors different from the mean velocity field. Eddy viscosity models relate the Reynolds stresses to the mean rate of strain tensor. This linear relationship between the two tensors forces the Reynolds stress tensor predicted by the eddy viscosity model to have the same eigenvectors as the mean rate of strain tensor. This is not true in real life turbulent flows like those with flow separation, streamline curvature, re-attachment, separation bubbles, over complex surfaces, etc. 

There are three broad approaches that have been used for using eigenvector perturbations:
\begin{enumerate}
    \item Eigenvector perturbations to extremal states of production: In this approach the focus is not on what misalignments between the RANS model's predicted eigen-directions and the true eigen-directions are likely but what misalignments are allowed under physics. To find this maximum misalignment state the turbulent production mechanism is used. There are two alignments between the mean rate of strain tensor and the predicted Reynolds stress tensor that maximise and minimise turbulence production. These extremal states are the states that the RANS model's predicted eigen-directions are perturbed till.

    \item Eigenvector perturbations using incremental rotations and Euler bases: This approach focuses not on what states of misalignment are physically possible but those which are likely. Usually a Machine Learning model is trained to predict the sequence of rotations that would align the RANS model's predicted eigen-directions and the true eigen-directions. The reference system used for these chained rotations is the Euler angles. The orientation of the high fidelity data's eigen-directions can be reached, starting from the orientation of the RANS model's predicted eigen-directions using a specific sequence of intrinsic rotations, whose magnitudes are the Euler angles of the target orientation. These Euler angles can be learned by the Machine Learning model.

    \item Eigenvector perturbations guided by physical differential equations: In this approach additional physics based differential equations are used to guide the eigenvector perturbations. A key approach is where the investigators use the Reynolds Stress Transport equations to find the eigenvector perturbations that are consistent with the eigenvalue perturbations \cite{thompson2016strategy, thompson2019eigenvector}. The eigenvalue perturbations are applied and this Reynolds stress tensor is used in the momentum equations of the Reynolds Stress Transport equations. The resultant Reynolds stress from the Reynolds Stress Transport equation solution is used to infer the correct eigenvector perturbations. Consequently the eigenvector perturbations are dependent upon the eigenvalue perturbations which is not completely physically consistent. 
\end{enumerate}

Investigators have used Machine Learning models to control the eigenvectors perturbations \cite{xiao2017random, wu2018physics}. To improve generalization of data driven approaches we need to incorporate physics based information into the Machine Learning model. In a recent study \cite{matha2023improved, matha2023physically}
have shown that the eigenvector perturbation without any constraints can and does lead to unrealizable Reynolds stress tensor values and un-physical Reynolds stress tensor dynamics. This also leads to lack of self consistency where the final dynamics of the eigenspace perturbation is different from what is expected based on an intuitive understanding. They produce and develop a set of necessary physics-based constraints leading to a realizable eigenvector perturbations.

\section{Turbulence kinetic energy perturbation}
The Turbulent Kinetic Energy perturbation represents the Reynolds stress tensor predicted by the turbulence model via its Spectral decomposition. Then it perturbs the amplitude of this Spectral decomposition, after which the Reynolds stress tensor corresponding to this perturbed Spectral decomposition is reconstituted in physical space. In broad terms, the Reynolds stress tensor can be represented as an ellipsoid. The turbulent kinetic energy perturbation changes the \textit{size} of this ellipsoid. 

The need for the turbulent kinetic energy perturbation arises from the inability of eddy viscosity based models to capture the physics of turbulence using a scalar isotropic eddy viscosity coefficient. In literature authors have found large variations in the optimal value of this coefficient for different flows and also across different domains of the same turbulent flow case. The single scalar value of the isotropic eddy viscosity coefficient can be seen as a ``best-fit" compromise. Owing to this compromise there are large epistemic uncertainties that are introduced in the eddy viscosity based model predictions. The turbulent kinetic perturbations enable variation in the value of the turbulent eddy viscosity coefficient: $\frac{k}{k^*} = \frac{C^*_{\mu}}{C_{\mu}}$, where $k$ is the turbulent kinetic energy, $C_{\mu}$ is the value of the coefficient of eddy viscosity and starred quantities are the perturbed variants.

Currently there are no purely physics based approaches for the perturbation of the turbulent kinetic energy. Some investigators have used Machine Learning models with approximate parameterizations to perturb the turbulent kinetic energy. For example in \cite{cremades2019reynolds} the perturbation of $k$ is applied using the parameter $\eta \geq 1$ that prescribes the limits of the turbulent kinetic energy perturbation. The maximum perturbed turbulent kinetic energy corresponds to $k^* = \eta k$ and the minimum to $k^* = k/\eta$. The parameter $\eta$ is learned via a Machine Learning model. 

The absence of an established methodology for the perturbation of the turbulent kinetic energy is a big limitation of the Eigenspace Perturbation Framework. From the point of view of minimising the discrepancy between the RANS model predictions and high fidelity data errors in the turbulent kinetic energy can increase prediction discrepancy by orders of magnitude more than those in the predicted eigenvalues or eigenvectors. The challenge in developing this turbulent kinetic energy methodology is to establish limits on the perturbed turbulent kinetic energy. The lower limit on the perturbed turbulent kinetic energy is clearly that it should be non-negative. But an upper limit on the turbulent kinetic energy for a general turbulent flow is not identified. After this the question is the manner in which to apply these perturbations. The perturbations can be multiplicative, $k^* = \eta k$, or they can be additive, $k^* = \delta k + k$. A study into the relative stability of these two approaches would be a big advance in the application of data driven approaches to the perturbation methods for uncertainty quantification of RANS models.

While we have discussed the three different types of perturbations (that is to the eigenvectors, eigenvalues and the turbulent kinetic energy) separately they are applied in conjunction with each other. Thus while realizability conditions on each type of perturbation may be necessary, they may not be sufficient in the general case when all the types of perturbations are applied. Thus it is important to develop realizability constraints for the general case where all three types of perturbations are being applied. Also while studies have focused on the maintenance of realizability for the RANS predictions of the Reynolds stresses they have ignored the fact that RANS predictions are themselves often unrealizable. The Eigenspace Perturbation Method maintains a realizable Reynolds stress tensor to remain realizable but it also maintains an unrealizable Reynolds stress tensor to remain unrealizable. The effects of the Eigenspace Perturbation Method based perturbations on the unrealizable predictions from the RANS models needs to be studied even if to just assess if the perturbations are making these points more unrealizable.

\section{Summary, Conclusions and Future Directions}
Turbulent flows are important in problems for engineering design and analysis. At present and for the foreseeable future, computational fluid dynamics studies of turbulent flows rely on turbulence models both in Reynolds Averaged Navier Stokes modeling and sub-grid scale  models in Large Eddy Simulations. Such turbulence model based simulations suffer for different sources of uncertainties. In safety critical applications of engineering design we need reliable estimates of these predictive uncertainties. We focus on such Uncertainty Quantification  for CFD simulations. We review recent advances in estimating different components of uncertainty, including aleatoric, numerical and epistemic. We elaborate upon the use of Machine Learning (ML) algorithms for estimating these uncertainties. Most importantly, we elaborate upon limitations in these approaches. We outline the need for a physics based framework for the perturbation of the turbulent kinetic energy. We also highlight the need for realizability conditions to ensure that the perturbation, of the eigenvalues, eigenvectors and the turbulent kinetic energy are physically consistent. These include both necessary and sufficient realizability conditions. We point out that while perturbations of the modeled Reynolds stress maintain realizable state of the Reynolds stress it is still unexplored about what effect they have on unrealizable predictions of the turbulence model itself. While there is many studies focusing on the use of Machine Learning models for uncertainty estimation there needs to be interpretability for these machine learning models. Adding physics knowledge in Machine Learning models for CFD UQ is also very important. Based on this analysis we highlight central questions that need to be addressed and advocate focused steps to redress these limitations.



 \bibliographystyle{elsarticle-num} 
 \bibliography{cas-refs}





\end{document}